# Particle Size-Dependent Onset of the Quantum Regime in Ideal Dimers of Gold Nanospheres

*Jesil Jose,[‡] Ludmilla Schumacher,[‡] Mandana Jalali,\* Jan Taro Svejda, Daniel Erni\* and Sebastian Schlücker\**

Physical Chemistry, Department of Chemistry and Center of Nanointegration Duisburg-Essen (CENIDE), University of Duisburg-Essen, 45141 Essen, Germany

General and Theoretical Electrical Engineering (ATE), Faculty of Engineering, University of Duisburg-Essen, and Center for Nanointegration Duisburg-Essen (CENIDE), 47048 Duisburg, Germany

[‡]Authors contributed equally.

**ABSTRACT**

We report on the nanoparticle-size-dependent onset of quantum tunneling of electrons across the sub-nanometer gaps in three different sizes (30, 50, and 80 nm) of highly uniform gold nanosphere dimers. For precision plasmonics, the gap distance is systematically controlled at the level of single C-C bonds via a series of alkanedithiol linkers ($C_2$-$C_{16}$). The corresponding single-particle scattering spectra reveal that for the larger dimers the onset of quantum effects occurs at larger gap distances: $C_6$ for 80 nm, $C_5$ for 50 nm, and $C_4$ for 30 nm dimers. 2D non-local and quantum-corrected model (QCM) calculations reveal the physical origin for this experimental observation: the lower curvature of the larger particles leads to a higher tunneling current due to a larger effective conductivity volume in the gap. Our results have possible implications in scenarios where precise geometrical control over plasmonics properties is crucial such as in hybrid (molecule-metal) and/or quantum plasmonic devices.

**INTRODUCTION**

In dimers of noble metal nanoparticles with inter-gap distances in the range of nanometers, the plasmons of the individual particles interact with each other to form new hybridized plasmon modes with energies specified by the classical plasmon hybridization model.[1-3] It is possible to fine tune the behavior of these coupled modes by altering the size and/or shape of the nanoparticles as well as the morphology and/or size of the gap between them.[4-8] These coupled plasmon modes can be efficiently excited optically in the Vis/NIR region and are associated with large cross sections and significantly enhanced local electric fields in the nanometer sized volumes between the nanoparticles (hot spot).[9] There have been multiple studies where these exceptional features of the plasmons have been utilized for a variety of applications, namely highly sensitive enhanced spectroscopies,[10] hot electron generation,[11] catalysis, solar energy harvesting materials,[12] and non-linear optics.[13]

Despite the fact that the standard plasmon hybridization model holds true for most applications, it predicts a rather unphysical result of an ever-increasing electric field enhancement in plasmonic nanostructures with sub-nm gap distances. Quantum phenomenon such as electron tunneling, electron scattering, electron spill-out, quantum size effects and spatial non-locality of material polarization are expected to play a significant role in the optical response of plasmonic systems like these.[14, 15] Recent theoretical models that accounts for these quantum phenomena suggest a different pattern in the evolution of coupled plasmon modes in plasmonic structures with sub-nm gaps.[16-18] The modified theory predicts a rise in conductance across the gap before the particles make physical contact,[19-25] reducing the charge buildup across the gap and limiting maximum field enhancement achievable in plasmonic systems. Subsequent experiments have validated these predictions, establishing the plausibility of the quantum theory of plasmon coupling.[14-18]

Accessing the quantum regime has its own set of unique obstacles mainly due to the difficulty in fabricating sub-nm gap structures. Previous studies indicates that the plasmon coupling in nanoparticle dimer systems depends critically on the sub-nm gap distance between the dimers, the morphology of the gap as well as both shape and size of the nanoparticle. As such, precise control over the structural characteristics of the nanostructure is critical to ensure precision and reproducibility. Because of the electron dispersion at the length scales involved, top-down methods such as electron beam lithography are deemed inadequate for fabricating sub-nm cavities.[26, 27] We rectify this influence of ill-defined gap morphology, polydispersity and variations in gap distance[21, 28] by using colloidally prepared ideal dimers of spheres with high structural homogeneity and alkanedithiol linkers of varying lengths. Thanks to such precise control over the geometrical parameters, we were able to track the evolution of the plasmon coupling in three distinct sizes of dimers as the gap distance shrank to sub-nm, allowing us to better understand the role of the size of nanostructures in the transition from classical to quantum regime.

**RESULTS AND DISCUSSION**

Our previous studies on dimers of spherical and quasi-spherical nanoparticles examined different factors that could influence plasmon coupling in dimers such as monodispersity, sphericity and the surface stabilizing agents.[7] Our findings indicate that the critical element influencing the plasmon coupling for a given nanoparticle size and gap distance is the gap morphology.[7, 28] However, a systematic study of the evolution of bonding dipolar plasmon (BDP) coupling band with respect to the monomer size and gap distance has not been performed till date.

Highly spherical gold nanoparticles (AuNS) of varying diameters (30.9 ± 3.3 nm, 50.7 ± 1.8 nm, and 79.5 ± 2.8 nm) were prepared using controlled seeded growth method and subsequent etching[29, 30] (see SI 1). The dimers were prepared in high yield by sequentially assembling these monomeric nanoparticles on a substrate.[28] Briefly, the first AuNS is made to adsorb on a glass substrate via electrostatic interactions. Alkanedithiols with varying number of carbon atoms (C16, C10, C8, C6, C5, C4, C3 and C2) were then added, forming a self-assembled monolayer (SAM) on the nanoparticle surface. In the final step, the second AuNS is linked to the first via the free-standing thiol group at the linker molecule's accessible end. Assuming an all-trans conformation of linker molecules in the gap, our choice of linker molecules gives us a theoretically estimated range from 2.25 nm (C16) down to 0.66 nm (C2). We hypothesize that each distance is an upper limit since the presence of one or more gauche conformations would lead to smaller gap distances in reality. (SI 2)

Figure 1a, b and c show the transmission electron microscope (TEM) images of the three different sizes of 30, 50 and 80 nm (color coded with green, red, and blue respectively) dimers exhibiting excellent structural uniformity. This high structural uniformity along with precisely controlled gap distances and well-defined gap morphology results in a remarkably homogenous and regular dark-field (DF) scattering response across the gap distances at the single particle level, allowing us to identify the higher-order plasmon coupling modes like the bonding quadrupolar plasmon (BQP) and bonding octupolar plasmon (BOP).[28] Optical characterization at single-particle level has an added advantage, as it allows us to exclude monomers and randomly formed oligomers from the sample.[7, 28]

Figures 1d, e, and f show representative single-particle DF spectra from ideal dimers of three different sizes (30, 50 and 80 nm). Eight different gap distances were evaluated within each class of ideal dimers. For a specific AuNS diameter and gap distance, spectra from multiple dimers (N = 20, on average) were collected and analyzed for reliable statistics. For each dimer

class, the average spectral position ($\lambda_{Max}$) of the BDP mode is obtained, and the distribution of the $\lambda_{Max}$ is found to be narrow across different AuNS diameter and distance. The average spectral position of the BDP mode is given on the top of each class. According to Fig 1, the larger ideal dimers have a redshifted bonding plasmon mode relative to the smaller ones for the same gap distance. The C8 linked 80 nm dimer have a redshifted BDP band at 778.5±4.7 nm compared to 50 nm dimer (721.3±5.1 nm) and 30 nm dimer (674.3±7.3 nm) linked with C8. Furthermore, as the gap distance is reduced the BDP band begins to redshift in tandem with an increase in strength of higher order plasmon modes, indicating an enhanced coupling between the individual plasmons of the constituent AuNSs.

The single-particle DF scattering spectra of ideal dimers with varying gap distances were condensed into a single plot to visualize the influence of decreasing gap distance on bonding dipolar plasmon in ideal dimers (Figure 2a, b, and c). The bonding dipolar plasmon is formed by the hybridization of the dipolar plasmons of the monomer spheres, which is characterized by the accumulation of opposing charges across the gap. Because of the intense coulomb interaction between these charges, this mode is extremely sensitive to gap distance. Even a minor drop in gap distance causes drastic redshifts in the resonance wavelength for very narrow gaps like the ones we have here. As the length of the linker chain is reduced, the BDP peak strongly redshifts for all three sizes of dimers. The BDP mode redshifts from 649.9 nm at C16 to a maximum of 697.8 nm at C4 for 30 nm ideal dimers (Figure 2a). For 50 nm ideal dimers (Figure 2b), the BDP plasmon mode exhibits a steady red shift from 673.9 nm at C16 to 746.8 nm at C5. As for the 80 nm ideal dimers Figure 2c, when the gap distance is reduced from C16 to C6, we see a redshift in BDP band from 704.9 nm to 800.2 nm. The position of BDP band at C4 (697.8 nm), C5 (746.8 nm), and C6 (800.2 nm) corresponds to the maximum plasmon coupling strength achievable in 30 nm, 50 nm, and 80 nm ideal dimer systems respectively.

In contrast with classical theory's prediction of an exponentially increasing redshift of the BDP band with decreasing gap distance, when the gap distance is lowered to sub-nm values with the shortest linker molecules, the strength of redshift gradually fades before the final maximum value is reached. Beyond this point of maximum strength in plasmon coupling (C4 (697.8 nm), C5 (746.8 nm), and C6 (800.2 nm) for 30 nm, 50 nm, and 80 nm dimers respectively), a new regime is revealed where the behavior of plasmon coupling diverges drastically from classical predictions. The coupled BDP mode has now been replaced by a charge transfer plasmon (CTP') mode which blueshifts with further reduction in gap distance. This CTP' mode is characterized by the presence of an electron tunneling current across the gap which effectively screens the localized charges across the gap, reducing the plasmonic coupling. The redshift to blueshift crossover in ideal dimers corresponds to a threshold point where the reduction in charge due to quantum tunneling starts to balance the near-field coulomb interaction.[17, 19, 20, 24, 31]

Additionally, in Figure 2a, at C2, we observe a low energy plasmon mode at around ~800 nm which is much weaker in intensity and considerably broadened compared to the BDP mode. This peak is assigned to a low energy charge transfer plasmon mode (CTP) across the gap.[19, 32] The CTP peak in the case of 30 nm dimers is observable due to its higher energy, whereas in the case of 50 nm and 80 nm dimers, these peaks would be of even lower energy and would appear deeper in the NIR region which is beyond the observation capabilities of our experimental apparatus. Also note that in Figure 2c for the linker molecule C2 there is an anomalous redshift of the BDP band from C3. This contrasts with the trend expected, since at C2 there should be an increased tunneling of electrons across the gap compared to that of C3. This anomalous shift is attributed to the fact that the short linker molecules like C2 are extremely prone to dimerization of the sulfur headgroups and could in principle form a longer chain at long incubations during the ligand exchange step in the assembly of ideal dimers.

From the spectral position of the plasmon coupling mode in ideal dimers, we identify three distinct interaction regimes between the individual plasmons of the nanospheres (Figure 3). In the ideal dimers with large gaps, we observe a trend in accordance with the classical predictions where the spectral position of the BDP coupling band red shifts as the gap size is reduced using shorter linkers (classical regime). When the gap size approaches smaller values, however, the shift deviates from the classical trend due to the emergence non-local effects and the magnitude of the red shift begins to decrease (non-local regime) finally reaching a maximum. Beyond this point of maximum redshift in ideal dimers, at sub nm gap sizes, the behavior is dominated by quantum tunneling of electrons across the gap which decreases the dipole-dipole (BDP) coupling, giving rise to a charge transfer plasmon mode (CTP') which blue shifts with decreasing gap distance.[19] In Figure 3, the data point which marks the transition to quantum realm is highlighted in red. In comparing the evolution of plasmon coupling in the gap of these three different plasmonic systems, the important observation to be noted is that the onset of non-classical effects and its associated blue shift of the plasmon coupling mode occurs at a larger gap distance in the larger dimer systems compared to the smaller ones. i.e., the onset of non-classical behavior of plasmon modes originates beyond C4 in the case of 30 nm dimers, beyond C5 in the case of 50 nm dimers and for 80 nm structures this onset is further increased to bigger gap distances of C6. We emphasize that the identification of such small differences – a single carbon atom (C4 vs. C5 vs. C6) corresponds to ca. 0.1-0.15 nm or 100-150 pm – is only possible via precision plasmonics originating from the high structural uniformity of our dimers at the single-particle level.

Associated classical finite difference time domain (FDTD) simulations have been performed for three different sizes of ideal dimer systems (SI-3). Simulations matches the experimental observations at larger gap distances, such as the redshifted BDP mode of the larger dimers relative to the smaller ones and the progressive redshift as the gap distance is reduced. It fails, however, to account for the blue shift of the plasmon coupling band and the emergence of charge transfer plasmon mode in 30 nm dimers due to tunneling of electrons across the potential barrier at extremely short separations.

To understand the origin of this nanoparticle size dependent onset of quantum regime in ideal nanodimers, further numerical simulations were performed. In nanodimer with sub-nanometer gap sizes, as the electron scattering from the particle's boundaries affect the mean free path of the conducting electrons and additionally quantum tunneling probability plays a role, a spectral shift and broadening of the bonding dipolar plasmon (BDP) is expected[33], which classical formalisms fail to account for (Figure 2 and Figure 3). For an accurate description of such dimer's nonlocality, due to the gap sizes which are smaller than the Fermi wavelength of the electron, together with the quantum tunneling probability due to the sub-nanometer gap sizes has to be taken into account. The nonlocality has been included within the model based on the hydrodynamic model[34], while the quantum tunneling probability is incorporated via the quantum-corrected model (QCM)[17, 18], which phenomenologically bridges the quantum and classic plasmonics. As a full 3D modelling of spherical dimers is computationally costly and even for large diameters not tractable, nanowires are modelled since the underlying mechanisms are the same. Proceeding from the QCM, the quantum tunneling probability is translated into an artificial static conductivity of the gap and then included in the model through defining a fictitious material for the gap (cf. Figure 4 a)) with a permittivity based on the Drude model[18]. In such permittivity, the damping term includes the static conductivity as shown in the equation (1).

$$\varepsilon_g(\omega, l) = \varepsilon_0(\omega) + (\varepsilon_m^d(\omega) - \varepsilon_0(\omega))\exp(\frac{-l}{l_d}) - \frac{\omega_P^2}{\omega(\omega + i\gamma_g(l))}$$
$$\gamma_g(l) = \frac{\omega_P^2}{4\pi\sigma_0(l)}; \qquad \sigma(\omega, l) = i\omega\frac{\varepsilon_g(\omega, l) - 1}{4\pi} \qquad (1)$$

The real part of such fictitious material permittivity as well as the fictitious material conductivity along the direction of the gap ($y$ axes in the Figure 4 a)) for the case of the 30 nm gold dimer with 0.2nm air gap are depicted in the Figure 4, which shows that the corresponding material properties are both spatially and frequency dispersive. Subsequently in the Figure 5, the gap conductivity is compared for different dimer diameters with same gap sizes, which is the minimum separation distance between the two nanowires. The graph indicates that the artificial conductivity is dependent on both the gap distance and the wall's curvature. The maximum value of the conductivity mainly depends on gap distance, while the spatial distribution of the conductivity depends on the wall's curvature. This results in a nanowire's radius-dependent effective conductivity volume and correspondingly predicts a nanowire's radius-dependent onset of the quantum tunneling.

The underlying dimers are simulated based on the classical, the hydrodynamic[2] and the quantum correct models[18] for various gap sizes and the bonding dimer plasmon modes are represented in the Figure 6. As shown in this figure, the quantum corrected model starts to deviate from the hydrodynamic model as the gap sizes becomes smaller, which indicates the onset of quantum tunneling current. This deviation starts at larger gap sizes when the dimer diameter is larger, namely at 0.52 nm gap size for the case of 80 nm dimer and 0.48 nm for the case of the 30 nm dimer, and hence concludes a size dependent onset of the quantum tunneling current within the gap which is also observed in the measurements (cf. Figure 3). In order to quantitatively determine this deviation, a new parameter namely the hydrodynamic conductivity per unit area of the gap is defined via the equation (2).

$$\sigma_{\mathrm{hd}}\left(\frac{\mathrm{S}}{\mathrm{m}}\right) = \frac{\int_{gap} \frac{|\mathbf{J}_{\mathrm{hd}}|}{|\mathbf{E}|} dV}{(1\mathrm{m}^2)} \qquad (2)$$

Such hydrodynamic conductivity ascertains how easily the quantum tunneling current can flow within the gap and is determined for various gap sizes at the corresponding bonding dimer plasmon of the underlying structures (Table 1). Results show that as the hydrodynamic conductivity of the gap is larger for the case of the larger dimers, the quantum tunneling current starts at large gap sizes which is in accordance with the observed size dependent onset of the quantum tunneling current.

**CONCLUSION**

We studied the evolution of the scattering spectra of individual ideal AuNP dimers of varying diameters and gap distances for identifying the onset of quantum effects. A statistical analysis of multiple single ideal dimers revealed that the gap distance at which the experimental observation starts to deviate from the classical predictions critically depends on the size of the nanospheres. An earlier onset of quantum effects at a larger gap distance in larger ideal dimers is observed (Figure 3). 2D numerical simulations incorporating the non-local effects (hydrodynamic model) and the quantum tunneling probability (quantum-corrected model) due to the sub-nm gap distance, establish that the onset of deviation of quantum-corrected model from hydrodynamic model at very small gap distances have a strong dependence on the diameter of the nanostructure. A new parameter called the hydrodynamic conductivity per unit area of the gap is defined to quantify this nanoparticle size-dependent onset of quantum tunneling.

This study provides a complete understanding of the factors that dictates the quantum response of a plasmonic system, an important contribution in developing next generation nano-antennas and sensors. Additionally, it would be interesting to extend these experiments for different nanostructures of alternative cavity morphologies such as cubes or rods where the gap conductivity profile would vary significantly from a sphere. The ability to tailor nanostructures with highly precise and reproducible geometrical features and additional control on manipulating their sub-nm cavities presents us with wide range of applications in fields ranging from molecular electronics to development of novel quantum plasmonic devices.

**ASSOCIATED CONTENT**

**Supporting Information**.

**AUTHOR INFORMATION**

**Corresponding Authors**

*Emails:

mandana.jalali@uni-due.de

daniel.erni@uni-due.de

sebastian.schluecker@uni-due.de

**Notes**

The authors declare no competing financial interest.

**ACKNOWLEDGMENT**

Financial support from the Deutsche Forschungsgemeinschaft (DFG, German Research Foundation; project number 278162697 – SFB 1242 Non-equilibrium dynamics of condensed matter in the time domain, project A04) and Evonik industries (PhD fellowship for LS) is acknowledged.

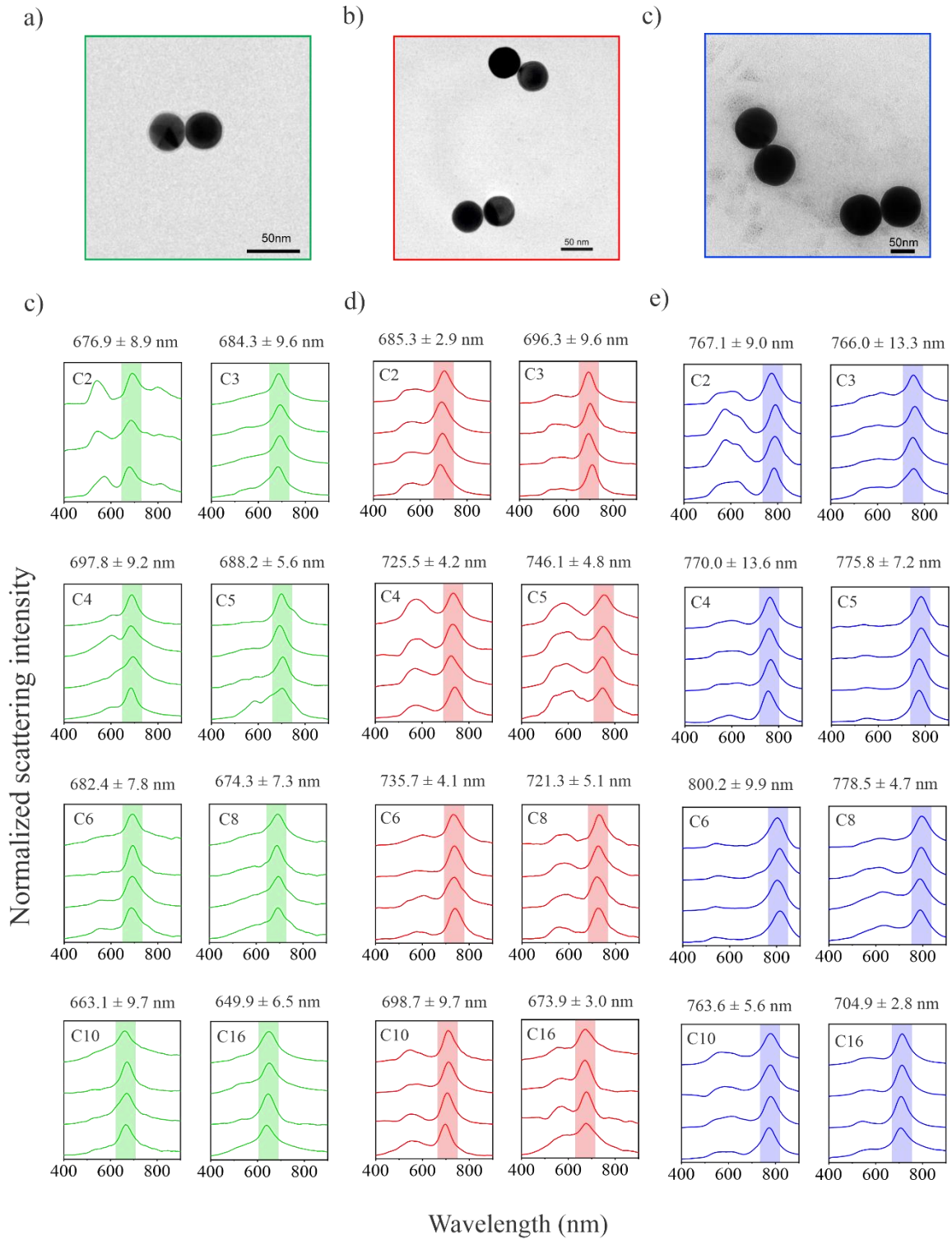

**Figure 1**: Transmission electron microscope images of different sizes of ideal dimers (30 nm, 50 nm, and 80 nm dimers are color coded green, red, and blue respectively) and the corresponding single particle dark field scattering spectra at different gap sizes. **a)** C8 linked 30 nm ideal dimer, **b)** C8 linked 50 nm ideal dimers, **c)** C8 linked 80 nm ideal dimers. **d), e), f)** Single particle dark field spectra of 30 nm dimers, 50 nm dimers and 80 nm dimers respectively, linked via alkanedithiol linkers with varying number of carbon atoms. The BDP band is highlighted, and the averaged band position is given at the top of each set of spectra.

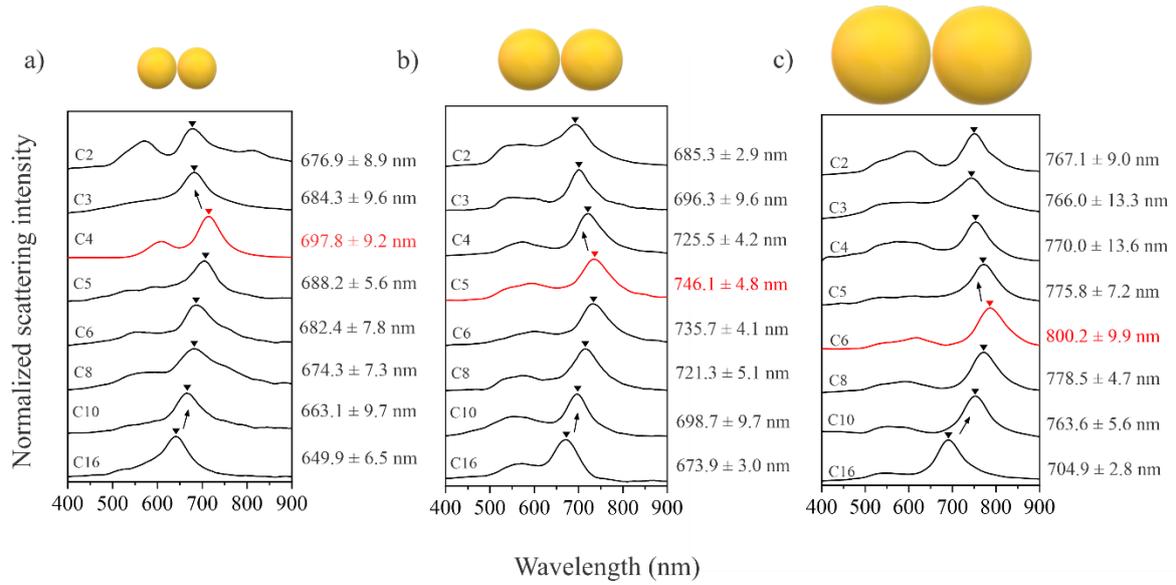

**Figure 2**: The experimental and computed elastic scattering spectra series of the three different ideal dimers. The spectra at the gap size beyond which a blue shift occurs is marked in red. The arrows guide the direction of the shift. Averaged position of the BDP band is given to the right of the spectra. **a)** 30 nm ideal dimers. The blue shift of the BDP band begins once the gap size is decreased beyond C4. The intensity of the BQP band is weak due to the smaller size of the spheres. At the lowest gap size of C2, a lower order CTP is visible at around 800 nm. **b)** 50 nm ideal dimers, and **c)** 80 nm ideal dimers. Notice that the blue shift of the BDP plasmon band, which is an indication of the onset of the non-classical phenomena, has an earlier onset in the bigger 80 nm dimers compared to the 50 nm and 30 nm dimers.

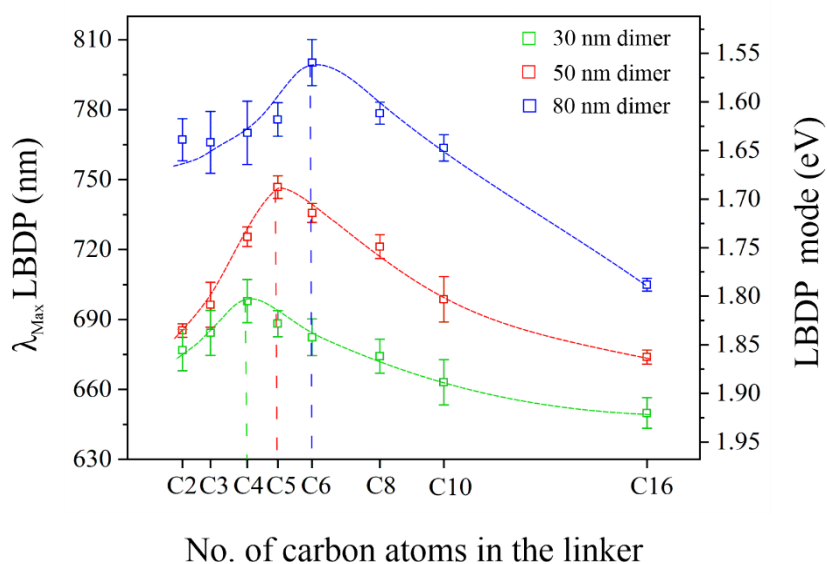

**Figure 3**: BDP band energy shift as a function of linker length, i.e. gap size. The earlier onset of the transition from the classical to the non-classical regime for larger dimers is indicated: C6 for 80 nm dimers, C5 for 50 nm dimers, and C4 for 30 nm dimers.

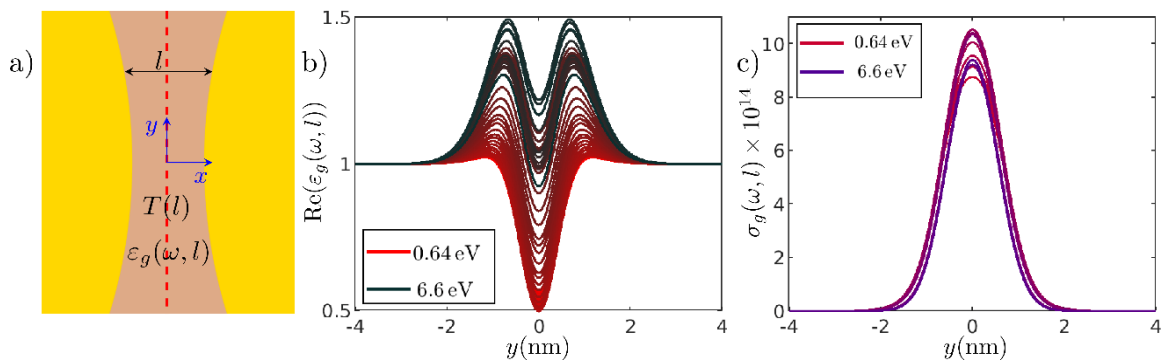

**Figure 4: a)** The schematic of the gap volume depending on the nanodimer geometry where the red dashed line extends along the direction of the gap ($y$ axes). **b)** The real part of the fictitious material permittivity for various energies along the $y$ axes, as well as **c)** the gap conductivity for various energies.

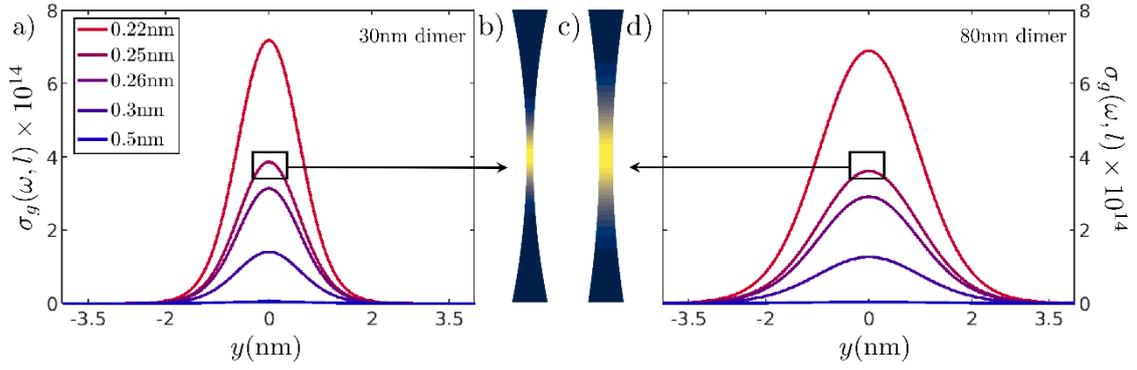

**Figure 5:** The conductivity inside the gap region for various gap sizes in both **a)** 30nm, and **d)** 80nm dimers along the $y$ axes together with the gap's conductivity profiles for **b)** 30nm and **c)** 80nm dimers.

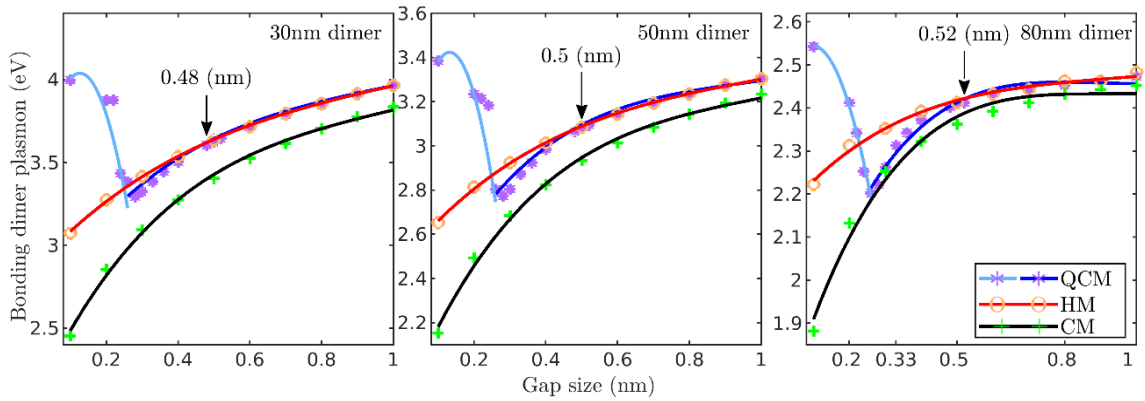

**Figure 6:** The bonding dimer plasmon in 30nm (left), 50nm (middle), and 80nm (right) dimers versus the gap sizes based on the quantum corrected model (QCM), the hydrodynamic model (HM) as well as the classical model (CM), where the markers are the simulation results which are then fitted to a polynomial (solid lines). The arrow in each graph indicates the onset of deviation between the quantum corrected and the hydrodynamic models.

|  |  | 30nm | 50nm | 80nm |
|---|---|---|---|---|
| BDP (eV) | 0.2nm | 3.88 | 3.23 | 2.41 |
|  | 0.3nm | 3.29 | 2.80 | 2.22 |
|  | 0.4nm | 3.44 | 2.98 | 2.34 |
|  | 0.5nm | 3.61 | 3.08 | 2.40 |
| $\sigma_{\text{hd}} \times 10^{-14}$ $(\frac{S}{m})$ | 0.2nm | 3.45 | 4.83 | 7.0 |
|  | 0.3nm | 0.48 ⟶ | 0.62 ⟶ | 0.78 ⟶ |
|  | 0.4nm | 0.005 | 0.007 | 0.008 |
|  | 0.5nm | 0.0005 | 0.0006 | 0.0008 |

**Table 1:** The gap hydrodynamic conductivity for 30nm, 50nm, and 80nm dimers with various gap sizes at each corresponding bonding dimer plasmon (BDP). The hydrodynamic conductivity profiles for the 0.3nm gap size is illustrated in the insets.

# *Supporting Information*

# Particle Size-Dependent Onset of the Quantum Regime in Ideal Dimers of Gold Nanospheres

*Jesil Jose,[‡] Ludmilla Schumacher,[‡] Mandana Jalali,\* Jan Taro Svejda, Daniel Erni\* and Sebastian Schlücker\**

Physical Chemistry, Department of Chemistry and Center of Nanointegration Duisburg-Essen (CENIDE), University of Duisburg-Essen, 45141 Essen, Germany

General and Theoretical Electrical Engineering (ATE), Faculty of Engineering, University of Duisburg-Essen, and Center for Nanointegration Duisburg-Essen (CENIDE), 47048 Duisburg, Germany

[‡]Authors contributed equally.

\*Emails:
mandana.jalali@uni-due.de
daniel.erni@uni-due.de
sebastian.schluecker@uni-due.de

**Materials**

Gold(III) chloride trihydrate (HAuCl$_4$·3H$_2$O, ≥ 99.9%, Aldrich), cetyltrimethylammonium bromide (CTAB, ≥ 96%, Fluka), cetyltrimethylammonium chloride (CTAC, > 95%, TCI), sodium borohydride (NaBH$_4$, 96%, Aldrich), ascorbic acid (AA, ≥ 99%, AppliChem), 1,2-Ethanedithiol (C2, 98%, Sigma-Aldrich), 1,3-Propanedithiol (C3, >97%, TCI Chemicals), 1,4-Butanedithiol (C4, >95%, TCI Chemicals), 1,5-Pentanedithiol (C5, 96%, Sigma-Aldrich), 1,6-Hexanedithiol (C6, 96%, Sigma-Aldrich), 1,8-Octanedithiol (C8, >97%, TCI Chemicals), 1,10-Decanedithiol (C10, >97%, TCI Chemicals), 1,16-Hexadecanedithiol (C16, 99%, Sigma-Aldrich), (11-mercaptoundecyl)-N,N,N-trimethylammonium bromide (MUTAB, ≥ 90%, Aldrich), sodium bromide (NaBr, ≥ 99.5%, Aldrich), ethanol (EtOH, HPLC grade, Fisher Scientific), and acetonitrile (MeCN, HPLC grade, Fisher Scientific). All chemicals were used as received. Deionized water (resistivity of 18 MΩ·cm) was prepared using a Millipore Milli-Q system.

**Nanoparticle synthesis**

The monomeric building blocks of AuNS dimers of varying mean diameters of 30, 50 and 80 nm are synthesized using the seeded growth method and subsequent etching.[1,2] In short, small clusters of gold were formed by reducing Au$^{3+}$ ions with NaBH$_4$ in CTAB solution. These clusters were subsequently grown into small seeds of mean size ~23 nm by further addition of Au$^{3+}$ ions and ascorbic acid in CTAB solution. In the third step these seeds were anisotropically grown into large polyhedrons of three different sizes (~32 nm, 52 nm and 82 nm respectively, as precursors to the intended final sizes) by the addition of Au$^{3+}$ ions and ascorbic acid in CTAC solution. The amount of seeds added in the solution critically determines the final size of the polyhedrons. To yield a higher size of polyhedron, a smaller amount of seeds were added and for a smaller size higher amount of seeds were added. In the final step, Au$^{3+}$ ions are added as an etching agent to smoothen out the surface of the polyhedra to yield spheres with highly smooth surfaces.

**Dimer assembly**

Ideal dimers of super-spherical nanoparticles were prepared by a controlled assembly of monomeric AuNSs on a substrate.[3, 4] The AuNSs are referred to as the first and second particle according to the order in which they were added to the system. The glass substrate onto which the dimers are absorbed were thoroughly cleaned and were given a negative charge by washing them in RBS detergent solution. In the initial step, the first AuNS is made to attach to the glass substrate via the electrostatic interaction between the substrate and the positively charged CTAB bilayer surrounding the nanoparticle. The dithiol linkers were added in the second step along with a solution of NaBr which disturbs the CTAB bilayer and aids in formation of self-assembled monolayers (SAM) of dithiol linkers within 1 hour. After the formation of the SAM, the second AuNS is added to the system and is covalently conjugated to the first nanoparticle via the remaining unreacted thiol moiety of the dithiol linkers. Subsequently, the ideal dimers are stabilized with (11-mercaptoundecyl)-N,N,N-trimethylammonium bromide (MUTAB) for preventing aggregation after dispersion. The fabricated ideal dimers are separated from the glass substrate and dispersed into the solution via sonication in the final step.

**Nanoparticle characterization**

The super-spherical monomers in suspension and dimer particles on glass substrate and in suspension were characterized using UV-Vis extinction spectroscopy using a UV-Vis absorption spectrometer (Jasco, V-630). The high monodispersity of the monomer nanospheres is apparent from the highly narrow FWHM of the extinction peaks. The formation of the dimers is confirmed via the newly formed (Longitudinal Bonding Dipolar Plasmon) LBDP coupling band. The high monodispersity of the monomers/dimers and the high yield and consistency of dimers are further confirmed via the Transmission Electron Microscope imaging (Jeol JEM-2200FS). Each dimer sample is subjected to UV-Vis absorption spectroscopy and TEM imaging prior to single-particle DF scattering measurements to make sure that the dimers are formed in adequate amount.

**Dark-field spectroscopy**

Single-particle dark-field scattering measurements were performed using a home built optical set-up comprising a white light source (Zeiss HAL 100), an inverted microscope (Nikon Eclipse Ti-S) and a portable UV/Vis spectrometer (Oceanview QEPro). White light is focused onto the sample using an oil-immersion dark-field condenser (Nikon oil, 1.43-1.20) and the scattered light from the sample is collected using a 100x air immersion objective lens (Nikon, 0.9 NA). A Piezo nano-positioning stage (E-727, Physik Instrumente (PI) GmbH & Co) is used to accurately move the sample in nano/micro scale. A pinhole is introduced at the relayed real plane to spatially filter the scattered signal from a single particle. A flip mirror is used in the collection path to divert the signals to a CCD camera or the spectrometer for imaging and spectroscopy, respectively.

# UV-Vis extinction spectra

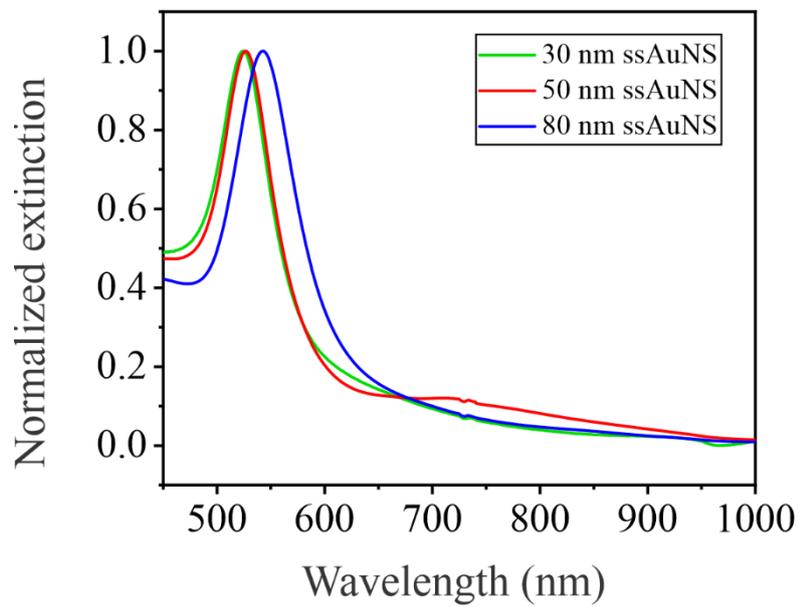

**SI 1**: Normalized UV-VIS extinction spectra of monomers of 30 nm, 50 nm, and 80 nm ideal spheres.

# High-Resolution TEM

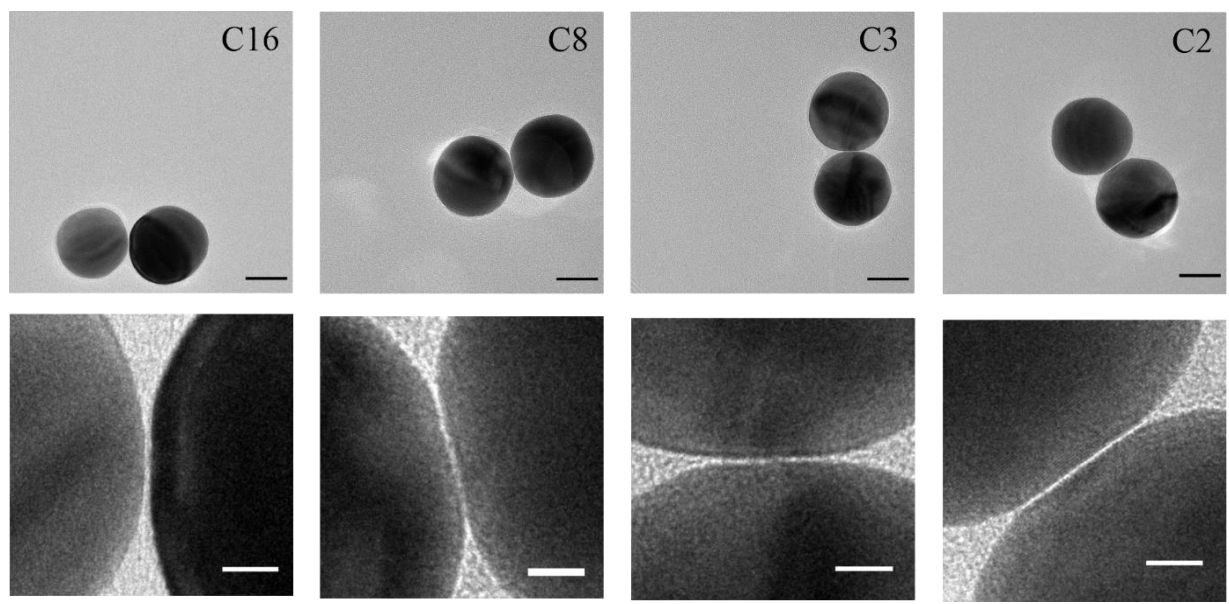

**SI 2:** Top panel: High resolution TEM images of 50 nm diameter ideal dimers linked with C16, C8, C3 and C2. Scale bar 25 nm. Bottom panel: Magnified view of the gap showing the progressive decrease in the gap distance as the number of carbon atoms in the linker molecule is decreased. Scale bar 5nm. The images are taken at an accelerating voltage of 120 kV and a magnification of 120,000x. The gap appears smaller due to parallax. A tilt series imaging to avoid parallax is difficult due to damage imparted to the gap by the electron beam under long exposure.

# FDTD simulations of scattering spectra

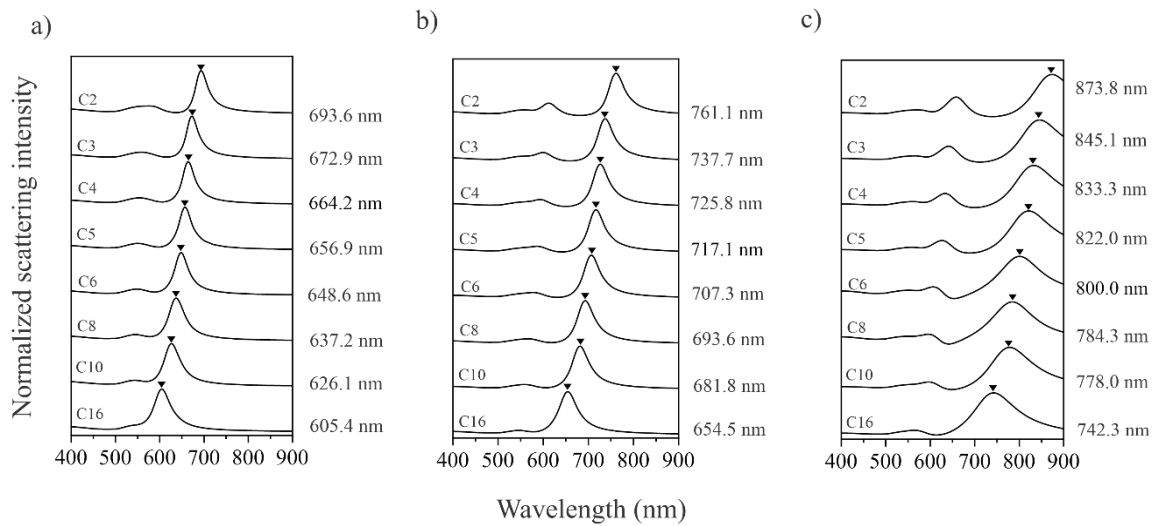

**SI 3:** The classical FDTD simulations of ideal dimers of gold nanospheres **a)** The computed elastic scattering spectra of the 30 nm ideal dimer. Notice that the higher order LBQP and LBOP modes are indistinguishable and very weak relative to the LBDP mode. **b)** The computed elastic scattering spectra of the 50 nm ideal dimer. **c)** The computed elastic scattering spectra of the 80 nm ideal dimer. The higher order plasmon peaks are intense and well separated. Also, the LBDP plasmon mode is much broader.

3D Finite Difference Time Domain (FDTD) software package from Lumerical solutions is used for simulating the scattering spectra of monomers and dimers. The dielectric function of gold is taken from the polynomial fitting of the experimental data obtained by Johnson and Christy. A mesh override region is defined around the gold nanoparticle using the non-uniform conformal variant mesh to ensure accuracy. The size of the mesh is defined in such a way that there are at least 4-5 mesh cells between the closest point (gap) between the dimers, to ensure accuracy of the results. A higher number of mesh cells was not feasible due to the associated increase in computational cost and time. The nanoparticles are illuminated using a linearly polarized total field scattered field (TFSF) plane wave source (400-900 nm). An analysis group placed in between the nanostructure and the TFSF source allows us to calculate the extinction cross section of the nanoparticle. At the same time another analysis group placed outside the TFSF source allows us to calculate the scattering cross section of the nanoparticle. The simulation region is terminated using perfectly matched layers (PMLs) to reduce the back-reflections from the boundaries into the simulation region.

Spherical dimers of diameters 30 nm, 50 nm and 80 nm were simulated with gap distances of 2.25 nm (C16), 1.5 nm (C10), 1.3 nm (C8), 1.1 nm (C6), 1 nm (C5), 0.9 nm (C4), 0.8 nm (C3) and 0.66 nm (C2). Even though the FDTD simulations fails to account for the blue shift of the plasmon coupling band at narrow gap sizes, it gives us some insights into the classical effects of plasmon coupling at larger gap distances; i) The 80 nm dimers show the highest red shift in Bonding dipolar plasmon (BDP) band due to their size, followed by 50 and 30 nm dimers, respectively ii) Larger dimers readily exhibit higher-order resonances compared to smaller dimers.

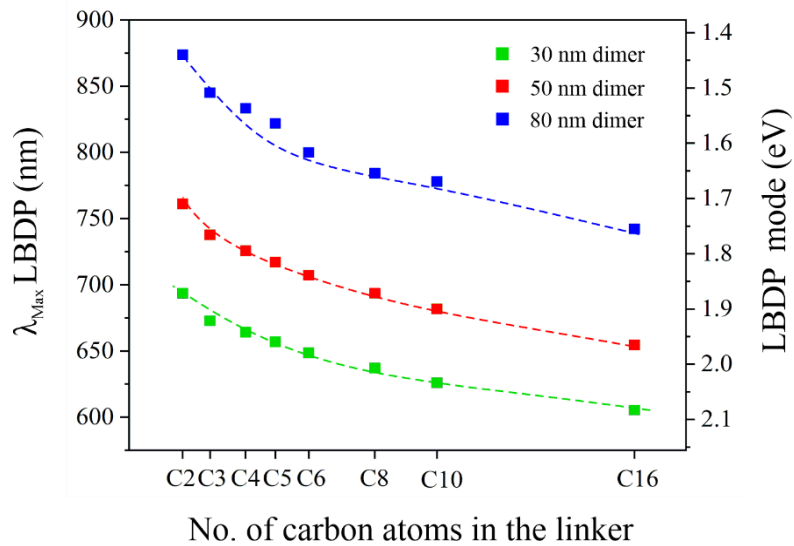

**SI 4:** The classical FDTD simulations of BDP band energy shift as a function of gap size.

# QCM simulations

The modelling is carried out with the classical Finite Element Method (FEM) solver COMSOL Multiphysics. The nonlocal material properties of the nanowires are described based on the nonlocal phasor polarization current $\mathbf{J}_{hd}$, which is coupled to the electric field based on the following relation[5].

$$[\nabla^2 + \frac{\omega(\omega + i\gamma)}{\beta^2}]\mathbf{J}_{hd}(\mathbf{r},\omega) = \frac{i\varepsilon_0 \omega_P^2 \omega}{\beta^2}\mathbf{E}(\mathbf{r},\omega)$$

in which the geometry dependent parameter $\beta$ is related to the Fermi velocity, and $\omega_P$ is the plasma frequency of gold. This equation is implemented as an auxiliary partial differential equation representing the material model of the nanowire domains, which is then coupled to the Maxwell equations, as an additional source term for the electric field. Within the gap, the hydrodynamic equation is further modified to contain the artificial static conductivity accounting for the quantum tunneling probability. The simulations are carried out for 30nm, 50nm, and 80nm dimers with gap sizes of 0.1nm, 0.2nm, 0.22nm, 0.24nm, 0.26nm, 0.28nm, 0.3nm, 0.33nm, 0.36nm, 0.4nm, 0.48nm, 0.5nm, 0.52nm, 0.6nm, 0.7nm, 0.8nm, 0.9nm, 1nm, 1.5nm, and 2nm in the 1eV to 6eV spectral range. To reduce the required computational resources and considering the symmetry of the dimer, the structure is cut into half (along the red dashed line in the Figure SI4), where the boundary condition is set to perfect electric conductor. Within the gap region, the nonlocal phasor polarization current follows the same boundary condition as a conventional displacement current. Such boundary condition set up assures that the structure is mirrored, while substantially reducing the simulation domain.

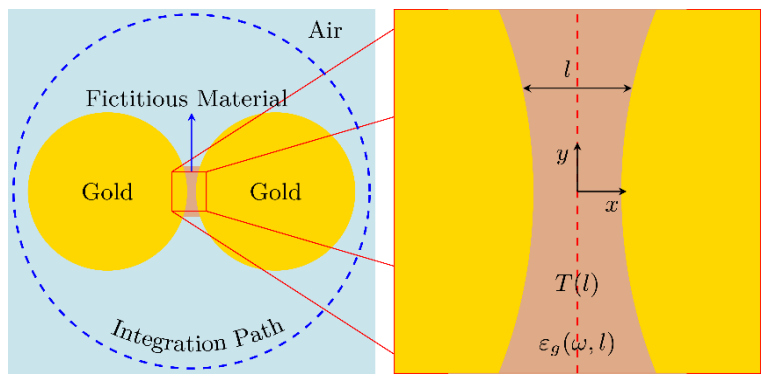

**SI 5:** The schematic of the simulation domain.

The simulation domain is meshed based on unstructured, non-uniform meshing, which resulted in about 25570 elements in the simulation of the smallest dimer with the smallest gap size, with the average mesh quality of 0.91. This simulation for all the gap sizes took 142h 25min 16s on an Intel(R) Xenon(R) Gold 6152 CPU @ 2.10GHz with 192GB Ram to solve.

The extinction cross section is calculated as the figure of merit, determining charge transfer plasmon and bonding dimer plasmon spectral positions.